\documentclass[fleqn,10pt]{wlscirep}
\usepackage[utf8]{inputenc}
\usepackage[T1]{fontenc}
\usepackage{amsmath,amssymb,physics,graphicx,bm,siunitx}
\usepackage{xcolor}
\usepackage{hyperref}

\title{Signatures of geostrophic turbulence in power spectra and third-order-structure function of offshore wind speed fluctuations}
\author[1]{So-Kumneth Sim}
\author[2]{Joachim Peinke}
\author[1,*]{Philipp Maass}
\affil[1]{Universit\"{a}t Osnabr\"{u}ck, Fachbereich Physik, Barbarastra{\ss}e 7, 49076 Osnabr\"uck, Germany}
\affil[2]{Universit\"at Oldenburg, {Institut f\"ur Physik \& ForWind}, K\"upkersweg 70, 26129 Oldenburg, Germany}
\affil[*]{maass@uos.de}

\newcommand{\rms}{\rm \scriptscriptstyle}

\begin{abstract} 
We analyze offshore wind speeds with a time resolution of one second
over a long period of 20 months for different heights above the sea
level.  Energy spectra extending over more than seven decades give a
comprehensive picture of wind fluctuations, including intermittency
effects at small length scales and synoptic weather phenomena at large
scales. The spectra $S(f)$ show a scaling behavior consistent with
three-dimensional turbulence at high frequencies $f$, followed by a
regime at lower frequencies, where $fS(f)$ varies weakly. Lowering the
frequency below a crossover frequency $f_{\rms 2D}$, a rapid rise of
$fS(f)$ occurs.  An analysis of the third-order structure function
$D_3(\tau)$ of wind speed differences for a given time lag $\tau$ 
shows a rapid change from negative to positive values of $D_3(\tau)$
at $\tau\simeq 1/f_{\rms 2D}$. Remarkably, after applying Taylor's
hypothesis locally, we find the third-order structure function to
exhibit a behavior very similar to that obtained previously from
aircraft measurements at much higher altitudes in the atmosphere. In
particular, the third-order structure function grows linearly with the
separation distance for negative $D_3$, and with the third power for
positive $D_3$.  This allows us to estimate energy and enstrophy
dissipation rates for offshore wind.  The crossover from negative to
positive values occurs at about the same separation distance of
400\,km as found from the aircraft measurements, suggesting that this
length is independent of the altitude in the atmosphere.
\end{abstract}

\begin{document}

\flushbottom
\maketitle
\thispagestyle{empty}

\section*{Introduction}
Understanding offshore wind properties is a central problem for
forecasting wind power and for estimating wind farm power outputs.
Due to the turbulent nature of wind flows in the atmosphere, this is a
challenging problem.  For three-dimensional (3D) homogeneous isotropic
turbulence, a description in terms of Kolmogorov's theory is
possible. Its hallmark is a scaling of kinetic energy
spectra with the wavenumber $k$ according to a $k^{-5/3}$ law (K41
scaling) \cite{Kolmogorov:1941c, Hunt/Vassilicos:1991}. This scaling
corresponds to a $f^{-5/3}$ scaling in the frequency domain when
applying Taylor's hypothesis \cite{Taylor:1938}.  Atmospheric
turbulence is, however, different because, apart from seasonal and diurnal
influences, scaling features are affected by geometric constraints
\cite{Wyngaard:2010}.  An improved understanding of its behavior is
one of the grand challenges in wind energy science
\cite{Veers/etal:2019}. For offshore wind, where 
obstacles such as buildings, trees, and mountains are absent,
one could ask whether a  generic characterization of wind speed fluctuations 
over many orders of time or frequency scales is possible.

Spectra of horizontal wind speeds $v=(v_x^2+v_y^2)^{1/2}$, with $v_x$
and $v_y$ being the components parallel to the Earth's surface, show a
deviation from K41 scaling. When a measurement at a small height $h$ in
the boundary layer is performed, an isotropic 
and homogeneous inertial (IHI) range of 3D turbulence
can no longer be assumed for length scales larger than
$h$. In energy (power) spectra $S(f)$ of wind speeds, the corresponding
crossover frequency $f_{\rms IHI}\simeq \bar v_h/h$, with $\bar v_h$
the mean wind speed at height $h$, marks the onset of an intermediate regime 
$f_{\rms 2D}<f<f_{\rms IHI}$ at lower frequencies, where $fS(f)$ varies
weakly.  This regime is sometimes referred to as the
spectral-gap \cite{VanderHoven:1957, Fiedler/Panofsky:1970,
  Metzger/Holmes:2008} and its features have been discussed
controversially. There is evidence that its properties are dependent
on the measurement height $h$ \cite{Larsen/etal:2016,
  Larsen/etal:2018}.  Several studies suggest that the spectrum in this
regime can show an $f^{-1}$ scaling
\cite{Calaf/etal:2013, Fitton:2013, Fitton/etal:2011,
  Drobinski/etal:2004} and different models have been developed to
explain such scaling \cite{Katul/etal:2012, Nickels/etal:2005,
  Hunt/Carlotti:2001, Nikora:1999, Korotkov:1976}.   
Other fitting
functions have been proposed also for describing the behavior
\cite{Cheynet/etal:2018, Larsen/etal:2019}.  

For a long time, it has also been debated 
whether atmospheric turbulence is characterized by scaling properties of
2D turbulence \cite{Kraichnan:1967,
  Gage:1979, Lilly:1989, Lindborg:1999, Danilov/Gurarie:2000}. For
isotropic 2D turbulence, the seminal paper by
Kraichnan~\cite{Kraichnan:1967} predicts a regime of $f^{-3}$ scaling
to occur at low frequencies as fingerprint of a forward enstrophy cascade,
followed by a $f^{-5/3}$ scaling at lower frequencies due to an
inverse energy cascade.  For geostrophic winds constrained by rotation
and stratification \cite{Callies/etal:2014, Oks/etal:2017}, the theory
by Charney \cite{Charney:1971} predicts that the potential enstrophy
is the relevant conserved quantity analogous to 2D turbulence.
Geostrophic turbulence behaves like 2D turbulence
\cite{Charney:1971,Vallgren/etal:2011} because of its forward
potential enstrophy cascade and conserved total energy
\cite{Vallgren/Lindborg:2010, Lindborg:2007}.  The theory of quasi-2D
geostrophic turbulence yields one regime of $f^{-3}$ scaling in energy
spectra.  Nevertheless, energy spectra obtained from aircraft
measurements show two scaling regimes with $f^{-5/3}$ and $f^{-3}$
scaling.  However, as pointed out by Lindborg \cite{Lindborg:1999},
their appearance is not in agreement with the theoretical prediction
for isotropic 2D turbulence, because the order of the regimes is
reversed.  This strongly suggests that the observed $f^{-5/3}$ scaling
regime is not due to 2D turbulence.  Stratified turbulence
\cite{Lilly:1983, Lindborg:2006} and cascades of inertia gravity waves
\cite{Dewan:1979, Callies/etal:2014} are commonly discussed as
possible explanations.
    
Here we show that spectra $S(f)$ of offshore wind speeds measured in
the North Sea exhibit the commonly observed main features for
frequencies $f>f_{\rms 2D}$ as discussed above.  For $f<f_{\rms 2D}$,
$S(f)$ rises strongly with decreasing $f$ and shows a behavior
consistent with the theoretical predictions for quasi-2D geostrophic
turbulence in an interval around $10^{-5}\,\si{Hz}$.  This interval,
however, is quite narrow and it is difficult to identify the $f^{-3}$
scaling clearly.

By studying the wind speed fluctuation in the time domain, we provide
further evidence that geostrophic turbulence dominates wind speed
fluctuations for $f<f_{\rms 2D}$.  This evidence comes from analyzing
third-order structure functions $D_3(\tau)$, i.e.\ the third moment
of differences between velocities separated by a time $\tau$.  The
function $D_3(\tau)$ changes sign from negative
\cite{Kolmogorov:1941c} to positive values at time lags $\tau\simeq
1/f_{\rms 2D}$, where a positive $D_3(\tau)$ indicates
a forward enstrophy cascade \cite{Lindborg:2007}.
The zero-crossing of $D_3(\tau)$ at $1/f_{\rms 2D}$ is remarkably sharp.
By revisiting spectra and third-order structure functions
obtained from aircraft measurements \cite{Nastrom/Gage:1983,
  Cho/Lindborg:2001}, we find that frequencies or wavenumbers
corresponding to $r_{\rms 2D}$ agree with corresponding crossover
frequencies to a $f^{-3}$ scaling regime.

\section*{Data set and data analysis}
\begin{figure*}[t]
\centering
\includegraphics[width=0.7\textwidth]{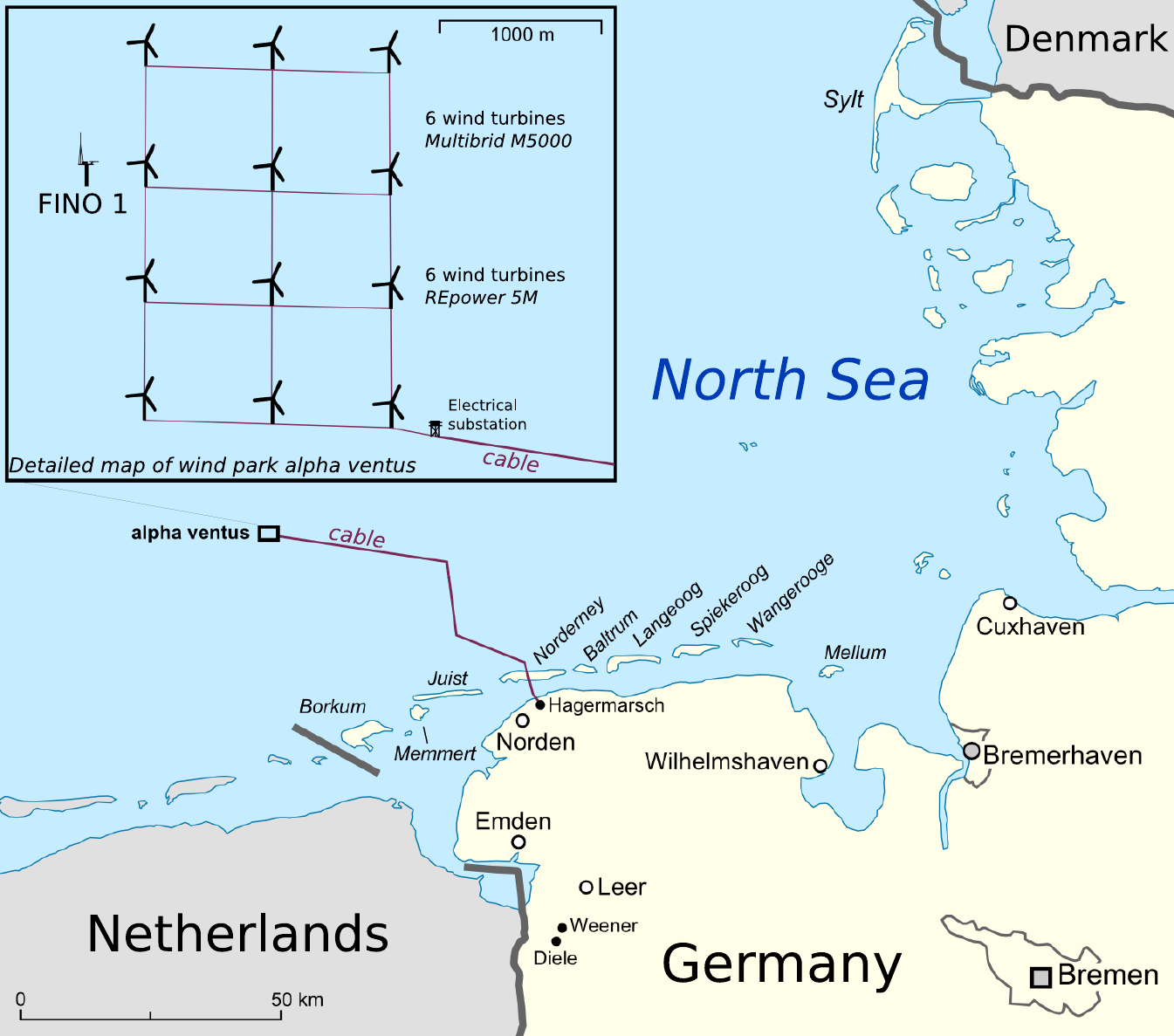}
\caption{FINO1 platform, located at Alpha Ventus wind farm at Borkum West in the North Sea
(54.3$^{\rm o}$~N--6.5$^{\rm o}$~W). 
Figure \href{https://de.wikipedia.org/wiki/Offshore-Windpark_alpha_ventus\#/media/Datei:Windpark_alpha_ventus_Lagekarte.png}{Wind park alpha ventus} 
adapted from Lencer (CC BY-SA 3.0).}
\label{fig:fino1}
\end{figure*}

Wind speeds were measured at the FINO1 platform in the North Sea,
which is located about 45 km north from the island Borkum
\cite{FINOdata}, see Fig.~\ref{fig:fino1}.  They were sampled by three-cup anemometers over
20~months, from September 2015 to April 2017, for eight different
heights $h$ between $30\,\si{m}$ and $100\,\si{m}$.  The time
resolution is $\Delta t=1\,\si{s}$, yielding time series with $N\cong
5\times10^7$ speed values for each height (for further details on the
data sampling and instrumentation, see
\href{https://www.bsh.de/EN/TOPICS/Monitoring_systems/MARNET_monitoring_network/FINO/fino_node.html;jsessionid=E124CD3AFECCADF31BD69BA0624BE5D3.live11292}{FINO
  - Database information}).

\begin{table}[t]     
\centering
\begin{tabular}{| l | l | l | l |}   
\hline
$h$                                          & $100\,\si{m}$       & $60\,\si{m}$           & $30\,\si{m}$          \\ \hline\hline
$\bar{v}_h$ $[\si{ms^{-1}}]$     & $9.2$                   & $8.6$                     & $8.2$                     \\ \hline
$\sigma_h$ $[\si{ms^{-1}}]$    & $4.8$                    & $4.6$                     & $4.3$                     \\ \hline
$F_{\rms NaN}$                         & $0.09\%$          & $0.09\%$          & $0.35\%$           \\ \hline
$\bar{T}_{\rms NaN}$             & $12\,\si{\minute}$  & $13\,\si{\minute}$  & $55\,\si{\minute}$ \rule[-0.3ex]{0ex}{3ex}   \\
\hline
\end{tabular}
\caption{Mean values $\bar v_h$ and standard deviations $\sigma_h$ of
  offshore wind speeds for three measurement heights at the FINO1
  platform in the North Sea.  The number $F_{\rms NaN}$ gives the
  fraction of NaN entries in the time series that remain after having
  interpolated single NaN entries. The time $\bar{T}_{\rms NaN}$ is
  the mean duration of intervals with successive NaN entries after
  single NaN interpolation.}
\label{tab:data-characteristics}   
\end{table}

The time series contain sequences of missing values of different
lengths.  These ``not a number'' (NaN) entries require special care in
the data analysis, in particular when calculating energy
spectra. Single missing values occur typically once a day, i.e.\ at
about every $10^5$th entry in the time series. A single NaN entry at a
time $t_{\rms NaN}$ has been replaced by the interpolated value
between the two wind speeds at the times $t_{\rms NaN}\pm\Delta t$.
In the resulting time series $v_t$ of wind speeds, the fraction
$F_{\rms NaN}$ of remaining NaN entries is given in
Table~\ref{tab:data-characteristics}. Time intervals with successive
NaN entries are typically much longer than one second, indicating a
temporary failure of the measurement device.  The mean duration $\bar
T_{\rms NaN}$ of the respective intervals is 12 minutes for the
measurement heights $h=60\,\si{m}$ and $100\,\si{m}$, and almost one
hour for $h=30\,\si{m}$, see Table~\ref{tab:data-characteristics}.
How we handle these longer time intervals of successive NaN entries is
explained below.

Diurnal variations of the offshore wind speeds did not show up as
significant patterns in spectra or structure functions and we
therefore did not apply a corresponding detrending of the data.  We
furthermore did neither consider seasonal variations nor changes of
meteorologic stability \cite{Cheynet/etal:2018}, because we expect
them to have only a weak effect on our principal results.  Seasonal
variations may affect our findings at very long times and
corresponding low frequencies only.

Results of our analysis are presented for the three measurements
heights $h=30\,\si{m}$, $60\,\si{m}$, and $100\,\si{m}$.  The mean
$\bar v_h$ and standard deviation $\sigma_h$ of the wind speeds for
these heights are listed in Table~\ref{tab:data-characteristics}.

\subsection*{Energy spectra}
For calculating energy spectra, we have used two methods to cope with
longer periods of missing values.
    
In the first method, we determined spectra $S_\alpha(f)$ separately
for all time intervals $\alpha$ with existing successive data. These
spectra were averaged in bins equally spaced on the logarithmic
frequency axis, yielding $S_{\rm ave}(f)$.

Specifically, let $\{v_n\}_\alpha=\{v_n^{(\alpha)}\,|\,
n=0,\ldots,N_\alpha-1\}$ be the $\alpha$th sequence of wind speeds
without NaN values, $\alpha=1,\ldots,N_{\rm seq}$, where $N_{\rm seq}$
is the number of these sequences. The discrete Fourier transform of
$\{v_n\}_\alpha$ is
\begin{equation}
\hat v_m^{(\alpha)}=\sum_{n=0}^{N_\alpha-1} v_n^{(\alpha)} e^{-2\pi i mn/N_\alpha}\,,\hspace{1em}
m=m_{\rms min}^{(\alpha)},m_{\rms min}^{(\alpha)}+1,\ldots,m_{\rms max}^{(\alpha)}\,,
\end{equation}
where $m_{\rms min}^{(\alpha)}=-\mathrm{int}((N_\alpha-1)/2)$ and
$m_{\rms max}^{(\alpha)}=\mathrm{int}(N_\alpha/2)$.  The energy
spectral density (``energy spectrum'')  of $\{v_t\}_\alpha$ at the frequency
$f_m^{(\alpha)}=m/T_\alpha$ with $T_\alpha=N_\alpha\Delta t$ is
\begin{equation}
S_m^{(\alpha)}=S_{-m}^{(\alpha)}=\frac{2\Delta t^2}{T_\alpha}\,|\hat v_m^{(\alpha)}|^2\,,
\hspace{1em} m=1,\ldots,m_{\rms max}^{(\alpha)}\,.
\end{equation}
These values $S_m^{(\alpha)}$ for frequencies $f_m^{(\alpha)}$ were
averaged in ten bins every decade with equidistant
spacing on a logarithmic frequency axis. The left and right border of
the $j$th bin are denoted as $f_j^-$ and $f_j^+$, respectively. The
averaged energy spectrum in the $j$th bin is
\begin{equation}
\bar S_j=\frac{\sum_{\alpha=1}^{N_{\rm seq}}\sum_{m=1}^{m_{\rm max}^{(\alpha)}}
S_m^{(\alpha)} I_j(f_m^{(\alpha)})}
{\sum_{\alpha=1}^{N_{\rm seq}}\sum_{m=1}^{m_{\rm max}^{(\alpha)}}
I_j(f_m^{(\alpha)})}\,,
\end{equation}
where $I_j(.)$ is the indicator function of the $j$th bin interval
$[f_j^-,f_j^+[$, i.e.\ $I_j(f)=1$ for $f\in[f_j^-,f_j^+[$ and zero
otherwise. The $\bar S_j$ value gives $S_{\rm ave}(f)$ at the
frequency $f=(f_j^-f_j^+)^{1/2}$,
\begin{equation}
S_{\rm ave}(f_j)=\bar S_j\,.
\label{eq:save}
\end{equation}
           
In the second method, each interval of successive missing values was
linearly interpolated between the two wind speed values terminating
the interval. The resulting series covers the total time span of
20~months and we calculated its energy spectrum $S_{\rm tot}(f)$.  This
spectrum should agree with $S_{\rm ave}(f)$ for frequencies $f\lesssim 1/\bar T_{\rms
  NaN}$ and perhaps higher frequencies. Indeed, as shown in
Fig.~\ref{fig:energy-spectra} below, the spectra $S_{\rm tot}(f)$ (full
circles) agree with $S_{\rm ave}(f)$ (open circles) in the
intermediate frequency range $10^{-4}\,\si{Hz}\lesssim f\lesssim
10^{-2}\,\si{Hz}$, and even up to frequencies of $10^{-1}\,\si{Hz}$ (not shown).
This demonstrates that $S_{\rm tot}(f)$ is reliable
for low frequencies $f<1/\bar T_{\rms NaN}$. 

\subsection*{Structure functions}
In the time domain, characteristic turbulence features can be
identified in the scaling behavior of structure functions.  The
structure function $D_q(\tau)$ of $q$th order at time lag $\tau$ is
the $q$th moment of the velocity fluctuation $[v_t - v_{t+\tau}]$,
\begin{equation}
D_q(\tau)=\left\langle [v_t - v_{t+\tau}]^q\right\rangle_t\,.
\label{eq:structf}
\end{equation}
Here, $\langle\ldots\rangle_t$ means an average over all times. 
We determined the structure functions without replacing missing values
by taking the average over all existing pairs $(v_t,v_{t+\tau})$. Knowing
$D_q(\tau)$, one can transform this to a function $D_q(r)$ with $r=\bar v_h\tau$,
where $\bar v_h$ is the mean wind speed averaged over the whole time series given in
Table~\ref{tab:data-characteristics}. This refers to applying Taylor's hypothesis ``globally''.

In a refined analysis, we take into account fluctuations of mean wind
speeds on the scale $\tau$. This corresponds to a method sometimes
referred to as local Taylor's hypothesis.  Specifically, for a given
pair of times $t$, $t+\tau$ we first calculated the average wind speed
$\bar v_{t,t+\tau}$ in the interval $[t,t+\tau[$, 
$\bar v_{t,t+\tau}=\sum_{\tau'=0}^{\tau-1} v_{t+\tau'}/\tau$. This
gives a distance $r_{t,t+\tau}=\bar v_{t,t+\tau}\tau$
corresponding to Taylor's hypothesis, i.e.\ a pair of values
$(r,\Delta v(r))=(r_{t,t+\tau},v_t-v_{t+\tau})$. The values
$\Delta v(r)^q$ are subsequently averaged in fifty bins every
decade with equidistant spacing on the logarithmic $r$ axis,
yielding $D_q^{\rm loc}(r)$, where the superscript indicates the
local use of Taylor's hypothesis. For comparison of 
$D_q^{\rm loc}(r)$ with $D_q(\tau)$, we can transform $D_q^{\rm loc}(r)$
back to a function depending on a time lag by using 
$D_q^{\rm loc}(\tau)=D_q^{\rm loc}(r/\bar v_h)$.
Differently speaking, applying the local Taylor’s hypothesis amounts to calculating
the right-hand side of 
Eq.~\eqref{eq:structf} for a transformed $\tau'=(\bar v_{t,t+\tau}/\bar v_h)\tau$.

In our analysis of the wind speed fluctuations in the time domain, we
focus on the structure function $D_3(\tau)$ and the kurtosis given by
\begin{equation}
\label{eq:kurtosis}
\kappa(\tau)=\frac{D_4(\tau)}{D_2(\tau)^2}\,.
\end{equation}
When using Taylor's hypothesis locally, we insert 
$D_2^{\rm loc}(\tau)$ and $D_4^{\rm loc}(\tau)$ in this equation, 
yielding $\kappa^{\rm loc}(\tau)$.

\begin{figure*}[t!]
\includegraphics[width=\textwidth]{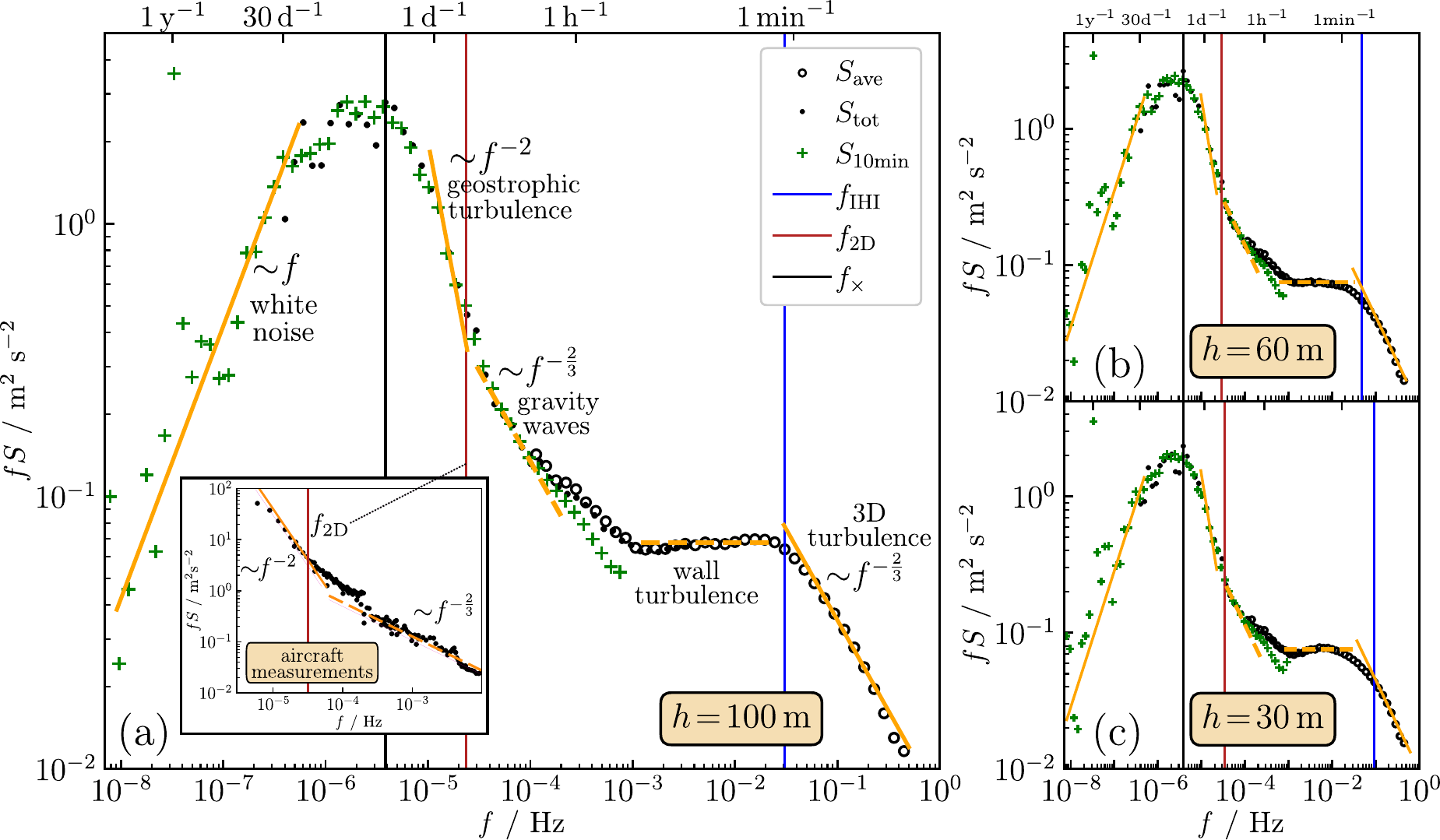}
\caption{Frequency-weighted energy spectra in a double-logarithmic
  representation for three different heights (\textbf{a})
  $h=100\,\si{m}$, (\textbf{b}) $h=60\,\si{m}$, and (\textbf{c})
  $h=30\,\si{m}$.  Open circles refer to $fS_{\rm ave}(f)$, where
  $S_{\rm ave}(f)$ is obtained by an averaging over all spectra of
  sub-sequences without missing values (see the description of the
  energy spectra calculation).  Full circles refer to the total
  spectrum $S_{\rm tot}(f)$, where linearly interpolated wind speeds
  were taken in all intervals of missing data. Green crosses mark
  spectra calculated from ten minutes averaged wind speeds in the
  period January 2005 to July 2021.  The vertical lines separate the
  various regimes: the blue line at frequency $f_{\rms IHI}=\bar
  v_h/(3h)$ separates the scaling regime of 3D turbulence from the
  intermediate regime, the red line at frequency $f_{\rms 2D}$
  separates the intermediate regime from the regime of quasi-2D geostrophic 
  turbulence, and the black line at frequency $f_\times$ marks the onset of
  uncorrelated wind speed fluctuations (white noise behavior).  The
  theoretical scaling laws expected in the regimes of geostrophic and
  3D turbulence are indicated by orange lines, as well as the white
  noise behavior at very low frequencies. The inset in (a) shows
    energy spectra obtained from aircraft measurements 
 [extracted from Ref.~\cite{Nastrom/Gage:1983} 
and mapped to the frequency domain by applying Taylor's hypothesis with a 
mean wind speed $30\,\si{ms^{-1}}$.]}
 \label{fig:energy-spectra}      
 \end{figure*}

\begin{figure*}[t]
\centering
\includegraphics[width=0.9\textwidth]{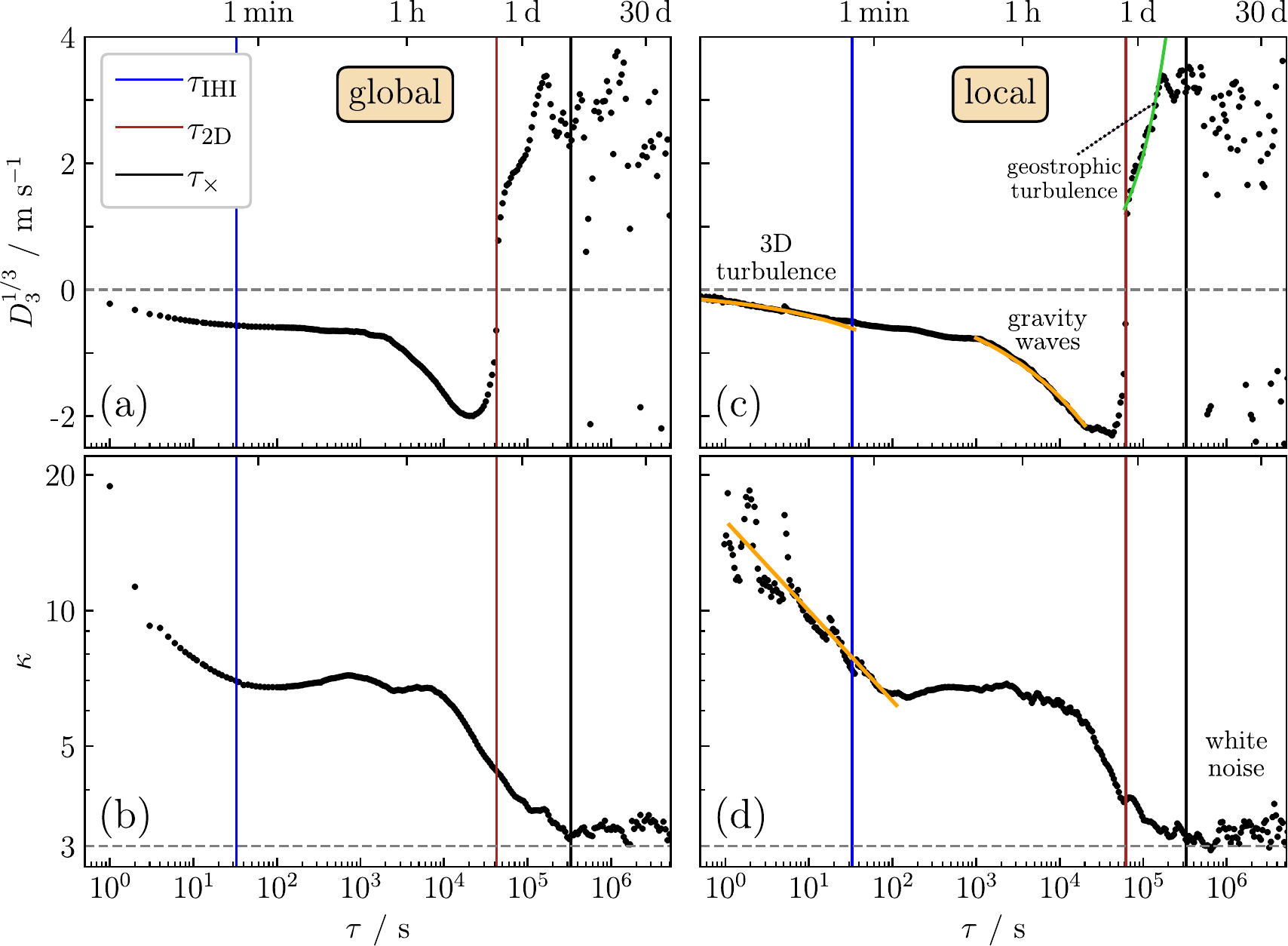}
\caption{(\textbf{a},\textbf{c}) Cubic root of the third-order
  structure function $D_3(\tau)$, and (\textbf{b},\textbf{d})
  kurtosis $\kappa(\tau)$ as a function of the time lag $\tau$ for the
  same measurement heights $h$ as in Fig.~\ref{fig:energy-spectra}. 
  Parts (\textbf{c}) and (\textbf{d}) show the results for $D_3^{\rm loc}(\tau)^{1/3}$ and $\kappa^{\rm
  loc}(\tau)$, when applying the local Taylor's hypothesis (see description of the data analysis).
  Vertical blue, red, and black lines correspond to the crossover
  frequencies in Fig.~\ref{fig:energy-spectra}.  In
  (\textbf{a},\textbf{b}), the sharp change from negative to
  positive values at $\tau\gtrsim1/f_{\rms 2D}$ indicates the
  transition to quasi-2D geostrophic turbulence. 
  In (\textbf{c}), the orange line in the short-time (IHI) regime of 3D turbulence marks the function
  $-(4/5)\epsilon_{\rms IHI}\bar v_h\tau$ with $\epsilon_{\rms IHI}=3\times10^{-3}\si{m^2s^{-3}}$ [cf.\ Eq.~\eqref{eq:D3-IHI}],
  and the orange line in the regime of turbulence induced by gravity waves marks the function
  $-2\epsilon\bar v_h\tau$ with $\epsilon=2.5\times10^{-5}\,\si{\meter\squared\per\cubic\second}$
  [cf.\ Eq.~\eqref{eq:D3-coriolis}]. 
  In (\textbf{d}), the orange line with slope (-0.2)
  indicates corrections to K41 scaling corresponding to an
  intermittency factor $\mu=0.45$.   At large time lags
  $\tau>1/f_\times$, $\kappa(\tau)$ is close to three, corresponding
  to a Gaussian distribution of velocity fluctuations.}
\label{fig:d3-kurtosis}       
\end{figure*}
 
\section*{Results and Discussion}
Figure~\ref{fig:energy-spectra}(a) shows the frequency-weighted energy
spectrum $fS$ vs.\ $f$ for the measurement height $h=100\,\si{m}$ in a
double-logarithmic representation. When comparing the data in
Fig.~\ref{fig:energy-spectra}(a) with the corresponding
frequency-weighted energy spectra for the other measurement heights in
the range $h=30-90\,\si{m}$, we have found almost the same functional
behavior. This is demonstrated in Figs.~\ref{fig:energy-spectra}(b) and
(c), where we show the results for $h=60\,\si{m}$ and
$30\,\si{m}$. Similarly, the structure functions $D_q(\tau)$ in the
time domain are nearly independent of $h$.
 
Figures~\ref{fig:d3-kurtosis}(a) and (c) show the results for the
third-order structure function for $h=100\,\si{m}$. We have plotted
$D_3(\tau)^{1/3}$ in a semi-logarithmic representation to make changes
of the function for small values easier visible.  In
Fig.~\ref{fig:d3-kurtosis}(a), $D_3(\tau)^{1/3}$ is displayed (indicated by ``global''), and
$D_3^{\rm loc}(\tau)^{1/3}$ in Fig.~\ref{fig:d3-kurtosis}(c) (indicated by ``local'').  The
corresponding results for the kurtosis $\kappa(\tau)$ and $\kappa^{\rm
  loc}(\tau)$ are shown in Figs.~\ref{fig:d3-kurtosis}(b) and
(d). Overall, the results in Figs.~\ref{fig:d3-kurtosis}(a) and (b)
are similar to that of their counterparts in
Figs.~\ref{fig:d3-kurtosis}(c) and (d), although there are differences
in detail.

In the following, we first discuss our results for the energy spectra
and structure functions in subsections referring to different
frequency and respective time regimes. In a final subsection, we
compare our findings for the third-order structure function in the
crossover regime to quasi-2D geostrophic turbulence with literature
results obtained from aircraft measurements.

\subsection*{IHI regime of 3D isotropic turbulence}
Above a frequency 
\begin{equation}
f_{\rms IHI}\sim\frac{\bar v_h}{h}\,,
\label{eq:f3D}
\end{equation}
with $\bar v_h$ the mean wind speed [see
  Table~\ref{tab:data-characteristics}], we see in
Fig.~\ref{fig:energy-spectra}(a) the signature of 3D turbulence, i.e.,
a behavior consistent with the K41 scaling. The border $f_{\rms IHI}$
of this frequency regime is marked by the vertical blue lines in the
figure, and the K41 scaling behavior by the solid lines with slope
$(-2/3)$.  

For the structure functions, the theory of
isotropic 3D~turbulence \cite{Kolmogorov:1941c} predicts a negative
\begin{equation}
D_3(r)=-\frac{4}{5}\epsilon_{\rms IHI}r\,,
\label{eq:D3-IHI}
\end{equation}
where $\epsilon_{\rms IHI}$ is the dissipation rate in the isotropic homogeneous inertial range.
Taking into account the intermittency
corrections to K41 scaling \cite{Kolmogorov:1962}, the kurtosis should scale as
\begin{equation}
\kappa(\tau)\sim\tau^{-4\mu/9}\,,
\label{eq:kappa-theory}
\end{equation}
where $\mu$ is the
intermittency factor and quantifies the amplitude of the logarithmic
correction in the scaling of the energy dissipation rate with
$r$. Values of $\mu$ lie in the range 0.2-0.5
\cite{Sreenivasan/etal:1977, Arneodo/etal:1996, Vindel/Yague:2011}.

Both $D_3(\tau)$ and $D_3^{\rm loc}(\tau)$ in
Figs.~\ref{fig:d3-kurtosis}(a) and (c) are negative in the regime
$\tau\lesssim\tau_{\rms IHI}=1/f_{\rms IHI}$. 
For the kurtosis shown in Figs.~\ref{fig:d3-kurtosis}(b) and (d),
the time $\tau_{\rms IHI}$ marks a crossover time from a regime where
$\kappa(\tau)$ decreases to another regime where it is nearly
constant.  That $\kappa(\tau)$ is much larger than~3 for small
$\tau$ reflects fat non-Gaussian tails in the distribution of wind
speed fluctuations for short times \cite{Morales/etal:2012}.

As for the laws \eqref{eq:D3-IHI} and \eqref{eq:kappa-theory}, the data in Figs.~\ref{fig:d3-kurtosis}(b)
and (d) can be well fitted to the respective equations,
while this is not the case for the data in Figs.~\ref{fig:d3-kurtosis}(a)
and (c). This shows that applying the local Taylor's hypothesis is needed here.

When fitting $-(4/5)\epsilon_{\rms IHI}\bar v_h\tau$
to the data for $D_3^{\rm loc}(\tau)$ in Fig.~\ref{fig:d3-kurtosis}(b) (orange line), we find $\epsilon_{\rms IHI}=3\times10^{-3}\si{m^2s^{-3}}$
for the dissipation rate. This value compares well with results reported in other studies of turbulence in the 
atmospheric boundary layer \cite{Munoz-Esparza/etal:2018}.
When fitting Eq.~\eqref{eq:kappa-theory} to the data for $\kappa^{\rm loc}(\tau)$ in Fig.~\ref{fig:d3-kurtosis}(d) (orange line), 
we obtain a slope corresponding to $\mu=0.45$. 
Deviations from the respective line could be explained by the fact that
cup anemometers loose precision for time lags approaching one second.

\subsection*{Intermediate regime of negative third-order structure function}
When $f$ becomes smaller than $f_{\rms IHI}$,
Figs.~\ref{fig:energy-spectra}(a)-(c) shows an intermediate regime (IR)
where $fS(f)$ first varies weakly and the K41 scaling is absent. In this regime,
the third-order structure function remains negative, see
Figs.~\ref{fig:d3-kurtosis}(a) and (c).

On scales $10^{-3}\,\si{Hz}\lesssim f<f_{\rms IHI}$ in the IR, $fS(f)$
is almost constant, or, equivalently, $S(f)\sim f^{-1}$.  Also,
$D_3(\tau)$ and $\kappa(\tau)$ remain nearly constant in the
corresponding time interval. 
We believe that this behavior reflects
turbulent wind patterns strongly influenced by the Earth's surface,
similarly as those found in wall turbulence experiments
\cite{Chandran/etal:2017} for Reynolds numbers larger than $6\times
10^4$ and in atmospheric boundary layers \cite{Calaf/etal:2013,
  Drobinski/etal:2004}.  We therefore refer to the regime $10^{-3}\,\si{Hz}\lesssim f<f_{\rms IHI}$
  as that of ``wall turbulence'', see Fig.~\ref{fig:energy-spectra}(a) and denote the lower limit
  of this regime as $f_{\rms wt}$, i.e.\ $f_{\rms wt}\simeq10^{-3}\,\si{Hz}$.
  The scaling $S(f)\sim f^{-1}$ can be reasoned
when considering wall turbulence to be governed by attached eddies
\cite{Perry/Chong:1982}.  Several models have been discussed to
explain this scaling \cite{Katul/etal:2012, Nickels/etal:2005,
  Hunt/Carlotti:2001, Nikora:1999, Korotkov:1976}. An $f^{-1}$ scaling in the energy spectra
  corresponds to a logarithmic dependence of $D_2(\tau)$ on $\tau$ 
  \cite{Banerjee/Katul:2013}. The second-order structure function follows this logarithmic behavior approximately for
  times $1/f_{\rms IHI}\lesssim \tau\lesssim 10^3\,\si{s}$ (not shown), similarly as it has been found in near-surface
  atmospheric turbulence on land \cite{Ghannam/etal:2018}.

Below $f_{\rms wt}$ in the IR, $fS(f)$ increases with decreasing
$f$. The structure function $D_3(\tau)$ in the corresponding time
interval first decreases to larger negative values, and after passing
a minimum rapidly rises towards zero. Interestingly, similar features have been seen
in the analysis of wind speed data sampled by aircraft.
Energy spectra obtained from aircraft measurements are shown in the inset
of Fig.~\ref{fig:energy-spectra}(a). These data were 
extracted from Ref.~\cite{Nastrom/Gage:1983} for different wavenumbers and
mapped to the frequency domain by applying Taylor's hypothesis with a 
mean wind speed $30\,\si{ms^{-1}}$ typical for the stratosphere.
As will be discussed further below, the frequency range $f_{\rms 2D}<f<f_{\rms wt}$
is likely connected to turbulent behavior induced by gravity waves.

\subsection*{Transition to quasi-2D geostrophic turbulence}
The IR regime terminates at a time lag $\tau_{\rms 2D}$, above which
$D_3(\tau)$ becomes positive, see Figs.~\ref{fig:d3-kurtosis}(a) and
(c). We interpret $f_{\rms 2D}$ as the frequency, below which
quasi-2D geostrophic turbulence is governing wind speed fluctuations.
According to the theory of geostrophic turbulence \cite{Charney:1971},
a scaling $fS\sim f^{-2}$ is predicted due to a forward cascade of
potential enstrophy \cite{Vallgren/Lindborg:2010}, analogous to the
enstrophy cascade of ideal isotropic 2D turbulence
\cite{Kraichnan:1967}.  Indeed, Figs.~\ref{fig:energy-spectra}(a)-(c)
show a sudden rapid of $fS$ increase towards lower $f$ for $f\lesssim
f_{\rms 2D}$.  When $f$ is close to $f_{\rms 2D}$, the data approach a
line indicating the expected scaling law $fS\sim f^{-2}$.  However,
the spectral data alone do not provide convincing evidence for a
transition to 2D turbulence. This is due to the limited extent of the
frequency interval, where the data are consistent with the expected
scaling behavior.

Strikingly, the transition becomes very well identifiable in
Figs.~\ref{fig:d3-kurtosis}(a) and (c).  Third-order structure
functions of quasi-2D geostrophic turbulence \cite{Lindborg:2007} are
similar to those of 2D turbulence, which are positive in general
\cite{Cerbus/Chakraborty:2017, Xie/Buehler:2018}.  The third-order
structure functions $D_3(\tau)$ in Figs.~\ref{fig:d3-kurtosis}(a) and
(c) indeed display a very sharp transition from negative to positive
values at $\tau=\tau_{\rms 2D}\sim 1/f_{\rms 2D}$.

At the frequency $f=1/\si{day}$ one could have expected a peak to
occur due to diurnal variations. Such a peak has indeed been observed
in the early analysis of onshore wind data by Van der Hoven
\cite{VanderHoven:1957}. A diurnal peak does not to occur in
Figs.~\ref{fig:energy-spectra}(a)-(c).  We believe that this is because
of weaker diurnal temperature variations of oceans compared to land
masses. For identifying scaling laws of atmospheric turbulence, this is an advantage
as well as the absence of mountains or other heterogeneities on land
that can inject long-lived coherent structures.
              
\subsection*{Three-day peak and white noise behavior at low frequencies}
For $\tau\gtrsim\tau_\times$, $\kappa(\tau)$ in
Figs.~\ref{fig:d3-kurtosis}(b) and (d) reaches a value
$\kappa(\tau)\simeq3$, reflecting Gaussian distributed wind speed
fluctuations.  The time $\tau_\times$ has a value of about 3~days and
corresponds to a frequency $f_\times=1/\tau_\times$, where $fS(f)$ in
Figs.~\ref{fig:energy-spectra}(a)-(c) runs through a peak maximum.
This peak has been attributed to the motion of low and high pressure
areas with linear dimension of about $10^3\,\si{km}$
\cite{VanderHoven:1957}.  If we assume Taylor's hypothesis to hold
even at large time scales of order $\tau_\times$, the corresponding
spatial scale $r_\times=\bar v_h\tau_\times\simeq3\times10^3\,\si{km}$
agrees with this length scale of low and high pressure areas.

For $r\gtrsim r_\times$, wind speed fluctuations can be expected to
become uncorrelated.  Accordingly, the energy spectrum should become
constant for $f<f_\times$.  To test this expectation, one needs very
long time series to suppress numerical noise in the spectra. The FINO1
project \cite{FINOdata} also provides ten minutes averaged wind speeds
in the long period January 2005 until July 2021.  Taking these data,
we calculated energy spectra $S_{\rm 10min}(f)$ with the same method as
used for obtaining $S_{\rm tot}$.  The results are represented by the
green crosses in Figs.~\ref{fig:energy-spectra}a-c and agree with
$S_{\rm tot}$ and $S_{\rm ave}$ for frequencies 
below $f_{\rms 2D}$. In the low-frequency regime $f<f_\times$, they indeed show a
behavior $fS_{\rm 10min}\sim f$ of a white noise spectrum. The
particular high value of $S_{\rm 10min}$ at the frequency of 1/year
reflects the seasonal cycle of winds at the yearly time scale.

\subsection*{Comparison of third-order structure function at low altitude with results from aircraft measurements}
\begin{figure*}[t]
  \centering
  \includegraphics[width=\textwidth]{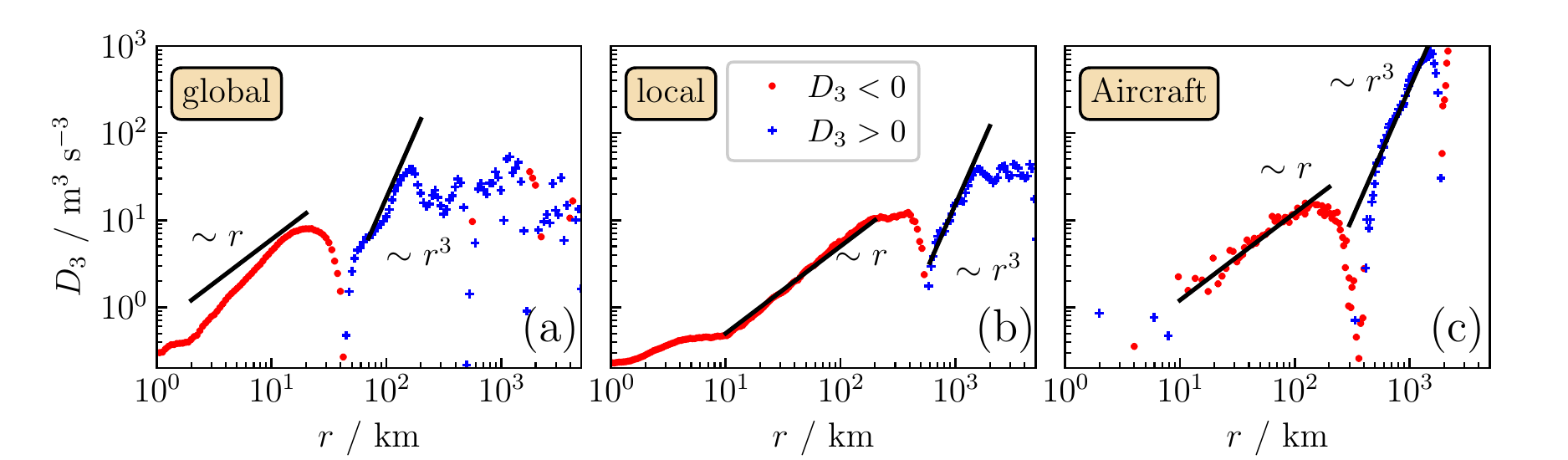}
  \caption{Third-order structure functions as function of separation
    distance $r$ for low altitudes by applying Taylor's hypothesis
    with (\textbf{a}) a globally averaged velocity, (\textbf{b}) with
    local averages of velocities, and (\textbf{c}) for high altitudes
    taken from aircraft measurements.  Blue crosses correspond to
    positive values of $D_3$ and red dots to negative ones.  The
    theoretical scaling laws expected in the regimes of geostrophic
    and 3D turbulence are indicated by black lines.}
  \label{fig:d3aircraft}
\end{figure*}

The third-order structure functions obtained from the
wind speeds measured at low altitudes of $\sim 100\,\si{m}$ above the
sea behave very similarly to those obtained from aircraft measurements
at very high altitudes of $\sim 10\,\si{km}$.  For this comparison, we
display our results for $D_3$ as a function of the distance $r$ in
Figs.~\ref{fig:d3aircraft} (a) and (b), where for transforming the
time lags $\tau$ to distances $r$, we used in (a) the mean wind speed
($r=\bar v_h\tau)$, and in (b) the local Taylor's hypothesis.  The
results from aircraft measurements were taken from Cho and Lindborg
\cite{Cho/Lindborg:2001} and are redrawn in
Fig.~\ref{fig:d3aircraft}(c).  The third-order structure functions in
Figs.~\ref{fig:d3aircraft}(a)-(c) show the same overall behavior: a
regime of negative $D_3$ at small $r_{\rms 2D}\lesssim\bar{v}_h
\tau_{\rms 2D}$ (red symbols) is followed by a regime of positive
values at large $r_{\rms 2D}\gtrsim\bar{v}_h \tau_{\rms 2D}$.

A merit of applying the local Taylor's hypothesis in our analysis
becomes clear when comparing the data in Figs.~\ref{fig:d3aircraft}(a)
and (b).  While in Fig.~\ref{fig:d3aircraft}(b) scaling regimes become
visible, this is not the case in Fig.~\ref{fig:d3aircraft}(a).
Notably, the results in Fig.~\ref{fig:d3aircraft}(b) show a linear
variation of $D_3$ with $r$ in the regime of negative $D_3$, and an
$r^3$-dependence in the regime of positive $D_3$, see the
corresponding lines in the figure. These lines were obtained by
least-square fits in the $r$ intervals $10\,\si{km}<r<200\,\si{km}$
and $600\,\si{km}<r<1500\,\si{km}$.

It is insightful to compare the energy dissipation rate $\epsilon$ and
the enstrophy flux $\eta$ at the different altitudes, which can be
extracted from the amplitude factors of the scaling laws. In the regime of linear variation of $D_3$
with $r$, the theory predicts, when incorporating Coriolis forces, \cite{Lindborg/Cho:2001}
\begin{equation}
D_3=-2\epsilon r.
\label{eq:D3-coriolis}
\end{equation}
Our analysis yields
$\epsilon=2.5\times10^{-5}\,\si{\meter\squared\per\cubic\second}$, which is of
similar magnitude as $\epsilon =
6\times10^{-5}\,\si{\meter\squared\per\cubic\second}$ obtained from
the aircraft data \cite{Lindborg/Cho:2001}.  For the forward enstrophy
cascade, the theory predicts\cite{Lindborg:1999, Lindborg/Cho:2001, Lindborg:2007} 
\begin{equation}
D_3(r)=\frac{1}{4}\eta r^3\,.
\label{eq:D3-enstrophy}
\end{equation}
Our analysis gives $\eta = 6\times10^{-17}\,\si{\per\cubic\second}$, which is about
20 times smaller than the value $\eta\simeq
1.5\times10^{-15}\,\si{s^{-3}}$ reported for the aircraft
measurements.

The length scale $r_{\rms 2D} \simeq 500\,\si{km}$, where $D_3$
crosses zero, can be estimated.  Geostrophic turbulence should arise
when rotation and stratification constrain synoptic-scale winds to be
nearly horizontal \cite{Callies/etal:2014, Oks/etal:2017}.  The length
scale at which rotation becomes as important as stratification is
described by the Rossby deformation radius with $r \simeq
500\,\si{km}$ as a standard estimation \cite{Lindborg:2006}, which is
the same as $r_{\rms 2D}$.  A dimensional analysis
\cite{Tung/Orlando:2003, Gkioulekas/Tung:2006, Vallgren/etal:2011}
that requires only the enstrophy flux $\eta$ and the energy
dissipation rate $\epsilon$ yields a further estimation of $r_{\rms
  2D}$.  Assuming $\eta r_{\rms 2D}^3 \sim \epsilon r_{\rms 2D}$, we
find $r_{\rms 2D}\sim\sqrt{\epsilon/\eta}\simeq 600\,\si{km}$ from the
data in Fig.~\ref{fig:d3aircraft}(b), and $\sqrt{\epsilon/\eta}\simeq
200\,\si{km}$ from the aircraft measurements in
Fig.~\ref{fig:d3aircraft}(c).  These estimates are of the same order
of magnitude.

In the analysis of the aircraft measurements, the regime of linear
variation $D_3(r)\sim r$ is related to a scaling
regime $S\sim k^{-5/3}$ of corresponding wavenumbers $k$ in kinetic
energy spectra \cite{Nastrom/etal:1984}.  Gravity waves are commonly
believed to be the physical mechanism leading to the corresponding
scaling behaviors with the same functional form as for 3D isotropic
turbulence \cite{Nastrom/etal:1984, Lindborg:1999, Cho/Lindborg:2001,
  Lindborg:2006, Callies/etal:2014}.  We can ask whether the energy
spectra for the wind speeds measured at low altitudes reflect this
finding.  The frequency interval corresponding to the $r$ regime
$10\,\si{km}<r<100\,\si{km}$ is $10^{-4}\,\si{Hz} < f <
10^{-3}\,\si{Hz}$.  In this regime, the local slopes in the
double-logarithmic plots in Figs.~\ref{fig:energy-spectra}(a)-(c)
indicate a behavior $fS\sim f^{-2/3}$ (or $S\sim f^{-5/3}$), shown by
the orange dashed line in Fig.~\ref{fig:energy-spectra}(a).

\section*{Conclusions}
Our analysis shows that the correlation behavior of offshore wind
speed fluctuations at times between a few hours and several days is in
agreement with the theory of quasi-2D geostrophic turbulence. While
features of this turbulence were seen in previous studies based on
aircraft measurements, we have found them here for low altitudes in
offshore wind. The third-order structure function in the time domain
shows a sharp transition from negative to positive values at a time
$\tau_{\rms 2D}$. Transforming the third-order structure function
$D_3$ from the temporal to the spatial domain, it is strikingly
similar to the aircraft data, if the local Taylor's hypothesis is used
for the transformation. In that case, both the linear variation with
the distance in a regime of negative $D_3$ (3D turbulence) and the
cubic variation with the distance in a regime of positive $D_3$ (2D
geostrophic turbulence) become visible. The transition between
negative and positive $D_3$ occurs at about the same length scale
$400\,\si{km}$ for the offshore wind at a height $100\,\si{m}$ and the
wind measured by aircraft at a height of about
$10\,\si{km}$. This strongly suggests that the length scale of the
transition to 2D geostrophic turbulence is independent of the
altitude.

We have given a comprehensive overview of the spectral behavior of
offshore winds covering times from seconds to years. At low
frequencies $f\ll f_\times\simeq 1/3\,{\rm days}$, a white noise
behavior is found, i.e.\ correlations between wind velocities are not
seen in the spectrum $S(f)$. Around $f_\times$, a peak appears in the
frequency-weighted spectrum $fS(f)$ that can be explained by the
motion of low- and high-pressure areas in the troposphere. For
$f>f_\times$, the spectral energy decreases with increasing
frequency. In a regime $f_\times<f<f_{\rms 2D}=1/\tau_{\rms 2D}$, it
decays as $S(f)\sim f^{-3}$ as predicted by the theory of geostrophic
turbulence.  For higher frequencies $f>f_{\rms 2D}$, results from
aircraft measurements \cite{Nastrom/Gage:1983, Nastrom/etal:1984} show
a weaker decay $S(f)\sim f^{-5/3}$, which has been interpreted as
resulting from 3D turbulence induced by gravity waves.  For the wind
measured at a low altitude $h\sim 100\,\si{m}$, we find indications of
such a regime for $f$ close to $f_{\rms 2D}$, but with increasing $f$
the weighted spectrum $fS(f)$ soon becomes flat, before it enters for
$f>f_{\rms IHI}$ a regime of 3D isotropic turbulence. The crossover
frequency $f_{\rms IHI}$ is about $\bar v/h$, where $\bar v$ is the mean
wind speed.  We believe that the intermediate regime 
$f_{\rms 2D}<f<f_{\rms IHI}$ has two parts: one at high frequency due to wall
turbulence with a behavior $S(f)\sim f^{-1}$, and a second one at
higher frequencies, which is influenced by gravity waves. A scaling
behavior according to gravity wave induced 3D turbulence, however,
becomes clearly visible only at higher altitudes.

Our findings shed new light onto the characterization of wind speed
fluctuations from micro- to synoptic scales and beyond. 
Frequencies of the order of $f_{\rms 2D}$ correspond to mesoscale processes on length scales of $10$-$100\,\si{km}$.
A better understanding of the relation between atmospheric phenomena on these mesoscales and microscales 
governing air flow around wind turbines and wind power plants, is considered as a grand challenge in wind energy 
science \cite{Veers/etal:2019}. This in particular concerns multiscale approaches, where a detailed simulation on 
microscales has to be connected to coarse-grained approaches on large scales. We believe that our findings on
geostrophic 2D turbulence below $f_{\rms 2D}$, the associated scaling of wind speed fluctuations, the
indications of gravity-wave induced 3D turbulence close to $f_{\rms 2D}$, and the overall characterization of the
different frequency regimes can improve the modeling of offshore wind flows across magnitudes of time scales.

\section*{Acknowledgements}
We thank M.~W\"achter for helping us with the data acquisition and the
BMWI (Bundesministerium für Wirtschaft und Energie) and the PTJ
(Projekttr\"ager J\"ulich) for providing the data of the offshore
measurements at the FINO1 platform. Financial support from the
Deutsche Forschungsgemeinschaft (MA 1636/9-1 and PE 478/16-1) is
gratefully acknowledged.


\begin{thebibliography}{10}
\expandafter\ifx\csname url\endcsname\relax
  \def\url#1{\texttt{#1}}\fi
\expandafter\ifx\csname urlprefix\endcsname\relax\def\urlprefix{URL }\fi
\providecommand{\bibinfo}[2]{#2}
\providecommand{\eprint}[2][]{\url{#2}}

\bibitem{Kolmogorov:1941c}
\bibinfo{author}{Kolmogorov, A.~N.}
\newblock \bibinfo{title}{Dissipation of energy in the locally isotropic
  turbulence}.
\newblock \emph{\bibinfo{journal}{Proc. R. Soc. London, Ser. A}}
  \textbf{\bibinfo{volume}{434}}, \bibinfo{pages}{15--17}
  (\bibinfo{year}{1991}).

\bibitem{Hunt/Vassilicos:1991}
\bibinfo{author}{Hunt, J. C.~R.} \& \bibinfo{author}{Vassilicos, J.~C.}
\newblock \bibinfo{title}{Kolmogorov's contributions to the physical and
  geometrical understanding of small-scale turbulence and recent developments}.
\newblock \emph{\bibinfo{journal}{Proc. R. Soc. London, Ser. A}}
  \textbf{\bibinfo{volume}{434}}, \bibinfo{pages}{183--210}
  (\bibinfo{year}{1991}).

\bibitem{Taylor:1938}
\bibinfo{author}{Taylor, G.~I.}
\newblock \bibinfo{title}{The spectrum of turbulence}.
\newblock \emph{\bibinfo{journal}{Proc. R. Soc. Lond. Ser. A}}
  \textbf{\bibinfo{volume}{164}}, \bibinfo{pages}{476--490}
  (\bibinfo{year}{1938}).

\bibitem{Wyngaard:2010}
\bibinfo{author}{Wyngaard, J.~C.}
\newblock \emph{\bibinfo{title}{Turbulence in the Atmosphere}}
  (\bibinfo{publisher}{Cambridge University Press}, \bibinfo{year}{2010}).

\bibitem{Veers/etal:2019}
\bibinfo{author}{Veers, P.} \emph{et~al.}
\newblock \bibinfo{title}{Grand challenges in the science of wind energy}.
\newblock \emph{\bibinfo{journal}{Science}} \textbf{\bibinfo{volume}{366}},
  \bibinfo{pages}{eaau2027} (\bibinfo{year}{2019}).

\bibitem{VanderHoven:1957}
\bibinfo{author}{Van~der Hoven, I.}
\newblock \bibinfo{title}{Power spectrum of horizontal wind speed in the
  frequency range from 0.0007 to 900 cycles per hour}.
\newblock \emph{\bibinfo{journal}{J. Atmos. Sci.}}
  \textbf{\bibinfo{volume}{14}}, \bibinfo{pages}{160--164}
  (\bibinfo{year}{1957}).

\bibitem{Fiedler/Panofsky:1970}
\bibinfo{author}{Fiedler, F.} \& \bibinfo{author}{Panofsky, H.~A.}
\newblock \bibinfo{title}{Atmospheric scales and spectral gaps}.
\newblock \emph{\bibinfo{journal}{Bull. Am. Meteorol. Soc.}}
  \textbf{\bibinfo{volume}{51}}, \bibinfo{pages}{1114 -- 1120}
  (\bibinfo{year}{1970}).

\bibitem{Metzger/Holmes:2008}
\bibinfo{author}{Metzger, M.} \& \bibinfo{author}{Holmes, H.}
\newblock \bibinfo{title}{Time scales in the unstable atmospheric surface
  layer}.
\newblock \emph{\bibinfo{journal}{Boundary-Layer Meteorol.}}
  \textbf{\bibinfo{volume}{126}}, \bibinfo{pages}{29--50}
  (\bibinfo{year}{2008}).

\bibitem{Larsen/etal:2016}
\bibinfo{author}{Lars{\'e}n, X.~G.}, \bibinfo{author}{Larsen, S.~E.} \&
  \bibinfo{author}{Petersen, E.~L.}
\newblock \bibinfo{title}{Full-scale spectrum of boundary-layer winds}.
\newblock \emph{\bibinfo{journal}{Boundary-Layer Meteorol.}}
  \textbf{\bibinfo{volume}{159}}, \bibinfo{pages}{349--371}
  (\bibinfo{year}{2016}).

\bibitem{Larsen/etal:2018}
\bibinfo{author}{Lars{\'e}n, X.~G.}, \bibinfo{author}{Petersen, E.~L.} \&
  \bibinfo{author}{Larsen, S.~E.}
\newblock \bibinfo{title}{Variation of boundary-layer wind spectra with
  height}.
\newblock \emph{\bibinfo{journal}{Q. J. R. Meteorolog. Soc.}}
  \textbf{\bibinfo{volume}{144}}, \bibinfo{pages}{2054--2066}
  (\bibinfo{year}{2018}).

\bibitem{Calaf/etal:2013}
\bibinfo{author}{Calaf, M.}, \bibinfo{author}{Hultmark, M.},
  \bibinfo{author}{Oldroyd, H.~J.}, \bibinfo{author}{Simeonov, V.} \&
  \bibinfo{author}{Parlange, M.~B.}
\newblock \bibinfo{title}{Coherent structures and the $k^{-1}$ spectral
  behaviour}.
\newblock \emph{\bibinfo{journal}{Phys. Fluids}} \textbf{\bibinfo{volume}{25}},
  \bibinfo{pages}{125107} (\bibinfo{year}{2013}).

\bibitem{Fitton:2013}
\bibinfo{author}{Fitton, G.}
\newblock \emph{\bibinfo{title}{Multifractal analysis and simulation of wind
  energy fluctuations (Analyse multifractale et simulation des fluctuations de
  l' \'energie \'eolienne)}}.
\newblock Ph.D. thesis, \bibinfo{school}{Universit\'e Paris-Est}
  (\bibinfo{year}{2013}).

\bibitem{Fitton/etal:2011}
\bibinfo{author}{Fitton, G.}, \bibinfo{author}{Tchiguirinskaia, I.},
  \bibinfo{author}{Schertzer, D.} \& \bibinfo{author}{Lovejoy, S.}
\newblock \bibinfo{title}{Scaling of turbulence in the atmospheric
  surface-layer: Which anisotropy?}
\newblock \emph{\bibinfo{journal}{J. Phys. Conf. Ser.}}
  \textbf{\bibinfo{volume}{318}}, \bibinfo{pages}{072008}
  (\bibinfo{year}{2011}).

\bibitem{Drobinski/etal:2004}
\bibinfo{author}{Drobinski, P.} \emph{et~al.}
\newblock \bibinfo{title}{The structure of the near-neutral atmospheric surface
  layer}.
\newblock \emph{\bibinfo{journal}{J. Atmos. Sci.}}
  \textbf{\bibinfo{volume}{61}}, \bibinfo{pages}{699 -- 714}
  (\bibinfo{year}{2004}).

\bibitem{Katul/etal:2012}
\bibinfo{author}{Katul, G.~G.}, \bibinfo{author}{Porporato, A.} \&
  \bibinfo{author}{Nikora, V.}
\newblock \bibinfo{title}{Existence of ${k}^{\ensuremath{-}1}$ power-law
  scaling in the equilibrium regions of wall-bounded turbulence explained by
  {H}eisenberg's eddy viscosity}.
\newblock \emph{\bibinfo{journal}{Phys. Rev. E}} \textbf{\bibinfo{volume}{86}},
  \bibinfo{pages}{066311} (\bibinfo{year}{2012}).

\bibitem{Nickels/etal:2005}
\bibinfo{author}{Nickels, T.~B.}, \bibinfo{author}{Marusic, I.},
  \bibinfo{author}{Hafez, S.} \& \bibinfo{author}{Chong, M.~S.}
\newblock \bibinfo{title}{Evidence of the ${k}_{1}^{\ensuremath{-}1}$ law in a
  high-{R}eynolds-number turbulent boundary layer}.
\newblock \emph{\bibinfo{journal}{Phys. Rev. Lett.}}
  \textbf{\bibinfo{volume}{95}}, \bibinfo{pages}{074501}
  (\bibinfo{year}{2005}).

\bibitem{Hunt/Carlotti:2001}
\bibinfo{author}{Hunt, J. C.~R.} \& \bibinfo{author}{Carlotti, P.}
\newblock \bibinfo{title}{Statistical structure at the wall of the high
  {R}eynolds number turbulent boundary layer}.
\newblock \emph{\bibinfo{journal}{Flow, Turbulence and Combustion}}
  \textbf{\bibinfo{volume}{66}}, \bibinfo{pages}{453--475}
  (\bibinfo{year}{2001}).

\bibitem{Nikora:1999}
\bibinfo{author}{Nikora, V.}
\newblock \bibinfo{title}{Origin of the ``$\ensuremath{-}1$'' spectral law in
  wall-bounded turbulence}.
\newblock \emph{\bibinfo{journal}{Phys. Rev. Lett.}}
  \textbf{\bibinfo{volume}{83}}, \bibinfo{pages}{734--736}
  (\bibinfo{year}{1999}).

\bibitem{Korotkov:1976}
\bibinfo{author}{Korotkov, B.~N.}
\newblock \bibinfo{title}{Kinds of local self-similarity of the velocity field
  of prewall turbulent flows}.
\newblock \emph{\bibinfo{journal}{Fluid Dyn.}} \textbf{\bibinfo{volume}{11}},
  \bibinfo{pages}{850--856} (\bibinfo{year}{1976}).

\bibitem{Cheynet/etal:2018}
\bibinfo{author}{Cheynet, E.}, \bibinfo{author}{Jakobsen, J.~B.} \&
  \bibinfo{author}{Reuder, J.}
\newblock \bibinfo{title}{Velocity spectra and coherence estimates in the
  marine atmospheric boundary layer}.
\newblock \emph{\bibinfo{journal}{Boundary-Layer Meteorol.}}
  \textbf{\bibinfo{volume}{169}}, \bibinfo{pages}{429--460}
  (\bibinfo{year}{2018}).

\bibitem{Larsen/etal:2019}
\bibinfo{author}{Lars{\'e}n, X.~G.}, \bibinfo{author}{Larsen, S.~E.},
  \bibinfo{author}{Petersen, E.~L.} \& \bibinfo{author}{Mikkelsen, T.~K.}
\newblock \bibinfo{title}{Turbulence characteristics of wind-speed fluctuations
  in the presence of open cells: A case study}.
\newblock \emph{\bibinfo{journal}{Boundary-Layer Meteorol.}}
  \textbf{\bibinfo{volume}{171}}, \bibinfo{pages}{191--212}
  (\bibinfo{year}{2019}).

\bibitem{Kraichnan:1967}
\bibinfo{author}{Kraichnan, R.~H.}
\newblock \bibinfo{title}{Inertial ranges in two‐dimensional turbulence}.
\newblock \emph{\bibinfo{journal}{Phys. Fluids}} \textbf{\bibinfo{volume}{10}},
  \bibinfo{pages}{1417--1423} (\bibinfo{year}{1967}).

\bibitem{Gage:1979}
\bibinfo{author}{Gage, K.~S.}
\newblock \bibinfo{title}{Evidence for a ${k}^{\ensuremath{-}5/3}$ law inertial
  range in mesoscale two-dimensional turbulence}.
\newblock \emph{\bibinfo{journal}{J. Atmos. Sci.}}
  \textbf{\bibinfo{volume}{36}}, \bibinfo{pages}{1950 -- 1954}
  (\bibinfo{year}{1979}).

\bibitem{Lilly:1989}
\bibinfo{author}{Lilly, D.~K.}
\newblock \bibinfo{title}{Two-dimensional turbulence generated by energy
  sources at two scales}.
\newblock \emph{\bibinfo{journal}{J. Atmos. Sci.}}
  \textbf{\bibinfo{volume}{46}}, \bibinfo{pages}{2026 -- 2030}
  (\bibinfo{year}{1989}).

\bibitem{Lindborg:1999}
\bibinfo{author}{Lindborg, E.}
\newblock \bibinfo{title}{Can the atmospheric kinetic energy spectrum be
  explained by two-dimensional turbulence?}
\newblock \emph{\bibinfo{journal}{J. Fluid Mech.}}
  \textbf{\bibinfo{volume}{388}}, \bibinfo{pages}{259--288}
  (\bibinfo{year}{1999}).

\bibitem{Danilov/Gurarie:2000}
\bibinfo{author}{Danilov, S.~D.} \& \bibinfo{author}{Gurarie, D.}
\newblock \bibinfo{title}{Quasi-two-dimensional turbulence}.
\newblock \emph{\bibinfo{journal}{Phys. Usp.}} \textbf{\bibinfo{volume}{43}},
  \bibinfo{pages}{863} (\bibinfo{year}{2000}).

\bibitem{Callies/etal:2014}
\bibinfo{author}{Callies, J.}, \bibinfo{author}{Ferrari, R.} \&
  \bibinfo{author}{B{\"u}hler, O.}
\newblock \bibinfo{title}{Transition from geostrophic turbulence to
  inertia-gravity waves in the atmospheric energy spectrum}.
\newblock \emph{\bibinfo{journal}{PNAS}} \textbf{\bibinfo{volume}{111}},
  \bibinfo{pages}{17033--17038} (\bibinfo{year}{2014}).

\bibitem{Oks/etal:2017}
\bibinfo{author}{Oks, D.}, \bibinfo{author}{Mininni, P.~D.},
  \bibinfo{author}{Marino, R.} \& \bibinfo{author}{Pouquet, A.}
\newblock \bibinfo{title}{Inverse cascades and resonant triads in rotating and
  stratified turbulence}.
\newblock \emph{\bibinfo{journal}{Phys. of Fluids}}
  \textbf{\bibinfo{volume}{29}}, \bibinfo{pages}{111109}
  (\bibinfo{year}{2017}).

\bibitem{Charney:1971}
\bibinfo{author}{Charney, J.~G.}
\newblock \bibinfo{title}{Geostrophic turbulence}.
\newblock \emph{\bibinfo{journal}{J. Atmos. Sci.}}
  \textbf{\bibinfo{volume}{28}}, \bibinfo{pages}{1087 -- 1095}
  (\bibinfo{year}{1971}).

\bibitem{Vallgren/etal:2011}
\bibinfo{author}{Vallgren, A.}, \bibinfo{author}{Deusebio, E.} \&
  \bibinfo{author}{Lindborg, E.}
\newblock \bibinfo{title}{Possible explanation of the atmospheric kinetic and
  potential energy spectra}.
\newblock \emph{\bibinfo{journal}{Phys. Rev. Lett.}}
  \textbf{\bibinfo{volume}{107}}, \bibinfo{pages}{268501}
  (\bibinfo{year}{2011}).

\bibitem{Vallgren/Lindborg:2010}
\bibinfo{author}{Vallgren, A.} \& \bibinfo{author}{Lindborg, E.}
\newblock \bibinfo{title}{Charney isotropy and equipartition in
  quasi-geostrophic turbulence}.
\newblock \emph{\bibinfo{journal}{J. Fluid Mech.}}
  \textbf{\bibinfo{volume}{656}}, \bibinfo{pages}{448--457}
  (\bibinfo{year}{2010}).

\bibitem{Lindborg:2007}
\bibinfo{author}{Lindborg, E.}
\newblock \bibinfo{title}{Third-order structure function relations for
  quasi-geostrophic turbulence}.
\newblock \emph{\bibinfo{journal}{J. Fluid Mech.}}
  \textbf{\bibinfo{volume}{572}}, \bibinfo{pages}{255--260}
  (\bibinfo{year}{2007}).

\bibitem{Lilly:1983}
\bibinfo{author}{Lilly, D.~K.}
\newblock \bibinfo{title}{Stratified turbulence and the mesoscale variability
  of the atmosphere}.
\newblock \emph{\bibinfo{journal}{J. Atmos. Sci.}}
  \textbf{\bibinfo{volume}{40}} (\bibinfo{year}{1983}).

\bibitem{Lindborg:2006}
\bibinfo{author}{Lindborg, E.}
\newblock \bibinfo{title}{The energy cascade in a strongly stratified fluid}.
\newblock \emph{\bibinfo{journal}{J. Fluid Mech.}}
  \textbf{\bibinfo{volume}{550}}, \bibinfo{pages}{207--242}
  (\bibinfo{year}{2006}).

\bibitem{Dewan:1979}
\bibinfo{author}{Dewan, E.~M.}
\newblock \bibinfo{title}{Stratospheric wave spectra resembling turbulence}.
\newblock \emph{\bibinfo{journal}{Science}} \textbf{\bibinfo{volume}{204}},
  \bibinfo{pages}{832--835} (\bibinfo{year}{1979}).

\bibitem{Nastrom/Gage:1983}
\bibinfo{author}{Nastrom, G.~D.} \& \bibinfo{author}{Gage, K.~S.}
\newblock \bibinfo{title}{A first look at wavenumber spectra from gasp data}.
\newblock \emph{\bibinfo{journal}{Tellus A}} \textbf{\bibinfo{volume}{35A}},
  \bibinfo{pages}{383--388} (\bibinfo{year}{1983}).

\bibitem{Cho/Lindborg:2001}
\bibinfo{author}{Cho, J. Y.~N.} \& \bibinfo{author}{Lindborg, E.}
\newblock \bibinfo{title}{Horizontal velocity structure functions in the upper
  troposphere and lower stratosphere: 1. observations}.
\newblock \emph{\bibinfo{journal}{J. Geophys. Res.: Atmos.}}
  \textbf{\bibinfo{volume}{106}}, \bibinfo{pages}{10223--10232}
  (\bibinfo{year}{2001}).

\bibitem{FINOdata}
\bibinfo{note}{{FINO1} project supported by the German Government through BMWi
  and PTJ. The database is accessible via \texttt{https://www.fino1.de/en}.}

\bibitem{Kolmogorov:1962}
\bibinfo{author}{Kolmogorov, A.~N.}
\newblock \bibinfo{title}{A refinement of previous hypotheses concerning the
  local structure of turbulence in a viscous incompressible fluid at high
  {R}eynolds number}.
\newblock \emph{\bibinfo{journal}{J. Fluid Mech.}}
  \textbf{\bibinfo{volume}{13}}, \bibinfo{pages}{82--85}
  (\bibinfo{year}{1962}).

\bibitem{Sreenivasan/etal:1977}
\bibinfo{author}{Sreenivasan, K.~R.}, \bibinfo{author}{Antonia, R.~A.} \&
  \bibinfo{author}{Danh, H.~Q.}
\newblock \bibinfo{title}{Temperature dissipation fluctuations in a turbulent
  boundary layer}.
\newblock \emph{\bibinfo{journal}{Phys. Fluids}} \textbf{\bibinfo{volume}{20}},
  \bibinfo{pages}{1238--1249} (\bibinfo{year}{1977}).

\bibitem{Arneodo/etal:1996}
\bibinfo{author}{Arneodo, A.} \emph{et~al.}
\newblock \bibinfo{title}{Structure functions in turbulence, in various flow
  configurations, at {R}eynolds number between 30 and 5000, using extended
  self-similarity}.
\newblock \emph{\bibinfo{journal}{Europhys. Lett. ({EPL})}}
  \textbf{\bibinfo{volume}{34}}, \bibinfo{pages}{411--416}
  (\bibinfo{year}{1996}).

\bibitem{Vindel/Yague:2011}
\bibinfo{author}{Vindel, J.~M.} \& \bibinfo{author}{Yag{\"u}e, C.}
\newblock \bibinfo{title}{Intermittency of turbulence in the atmospheric
  boundary layer: Scaling exponents and stratification influence}.
\newblock \emph{\bibinfo{journal}{Boundary-Layer Meteorol.}}
  \textbf{\bibinfo{volume}{140}}, \bibinfo{pages}{73--85}
  (\bibinfo{year}{2011}).

\bibitem{Morales/etal:2012}
\bibinfo{author}{Morales, A.}, \bibinfo{author}{W{\"a}chter, M.} \&
  \bibinfo{author}{Peinke, J.}
\newblock \bibinfo{title}{Characterization of wind turbulence by higher-order
  statistics}.
\newblock \emph{\bibinfo{journal}{Wind Energy}} \textbf{\bibinfo{volume}{15}},
  \bibinfo{pages}{391--406} (\bibinfo{year}{2012}).

\bibitem{Munoz-Esparza/etal:2018}
\bibinfo{author}{Mu{\~n}oz-Esparza, D.}, \bibinfo{author}{Sharman, R.~D.} \&
  \bibinfo{author}{Lundquist, J.~K.}
\newblock \bibinfo{title}{Turbulence dissipation rate in the atmospheric
  boundary layer: Observations and {WRF} mesoscale modeling during the {XPIA}
  field campaign}.
\newblock \emph{\bibinfo{journal}{Mon. Weather Rev.}}
  \textbf{\bibinfo{volume}{146}}, \bibinfo{pages}{351--371}
  (\bibinfo{year}{2018}).

\bibitem{Chandran/etal:2017}
\bibinfo{author}{Chandran, D.}, \bibinfo{author}{Baidya, R.},
  \bibinfo{author}{Monty, J.~P.} \& \bibinfo{author}{Marusic, I.}
\newblock \bibinfo{title}{Two-dimensional energy spectra in
  high-reynolds-number turbulent boundary layers}.
\newblock \emph{\bibinfo{journal}{J. Fluid Mech.}}
  \textbf{\bibinfo{volume}{826}}, \bibinfo{pages}{R1} (\bibinfo{year}{2017}).

\bibitem{Perry/Chong:1982}
\bibinfo{author}{Perry, A.~E.} \& \bibinfo{author}{Chong, M.~S.}
\newblock \bibinfo{title}{On the mechanism of wall turbulence}.
\newblock \emph{\bibinfo{journal}{J. Fluid Mech.}}
  \textbf{\bibinfo{volume}{119}}, \bibinfo{pages}{173--217}
  (\bibinfo{year}{1982}).

\bibitem{Banerjee/Katul:2013}
\bibinfo{author}{Banerjee, T.} \& \bibinfo{author}{Katul, G.~G.}
\newblock \bibinfo{title}{Logarithmic scaling in the longitudinal velocity
  variance explained by a spectral budget}.
\newblock \emph{\bibinfo{journal}{Phys. Fluids}} \textbf{\bibinfo{volume}{25}}
  (\bibinfo{year}{2013}).
\newblock \bibinfo{note}{125106}.

\bibitem{Ghannam/etal:2018}
\bibinfo{author}{Ghannam, K.}, \bibinfo{author}{Katul, G.~G.},
  \bibinfo{author}{Bou-Zeid, E.}, \bibinfo{author}{Gerken, T.} \&
  \bibinfo{author}{Chamecki, M.}
\newblock \bibinfo{title}{Scaling and similarity of the anisotropic coherent
  eddies in near-surface atmospheric turbulence}.
\newblock \emph{\bibinfo{journal}{J. Atmos. Sci.}}
  \textbf{\bibinfo{volume}{75}}, \bibinfo{pages}{943 -- 964}
  (\bibinfo{year}{2018}).

\bibitem{Cerbus/Chakraborty:2017}
\bibinfo{author}{Cerbus, R.~T.} \& \bibinfo{author}{Chakraborty, P.}
\newblock \bibinfo{title}{The third-order structure function in two dimensions:
  The {R}ashomon effect}.
\newblock \emph{\bibinfo{journal}{Phys. Fluids}} \textbf{\bibinfo{volume}{29}},
  \bibinfo{pages}{111110} (\bibinfo{year}{2017}).

\bibitem{Xie/Buehler:2018}
\bibinfo{author}{Xie, J.-H.} \& \bibinfo{author}{B{\"u}hler, O.}
\newblock \bibinfo{title}{Exact third-order structure functions for
  two-dimensional turbulence}.
\newblock \emph{\bibinfo{journal}{J. Fluid Mech.}}
  \textbf{\bibinfo{volume}{851}}, \bibinfo{pages}{672--686}
  (\bibinfo{year}{2018}).

\bibitem{Lindborg/Cho:2001}
\bibinfo{author}{Lindborg, E.} \& \bibinfo{author}{Cho, J. Y.~N.}
\newblock \bibinfo{title}{Horizontal velocity structure functions in the upper
  troposphere and lower stratosphere: 2. {T}heoretical considerations}.
\newblock \emph{\bibinfo{journal}{J. Geophys. Res.: Atmos.}}
  \textbf{\bibinfo{volume}{106}}, \bibinfo{pages}{10233--10241}
  (\bibinfo{year}{2001}).

\bibitem{Tung/Orlando:2003}
\bibinfo{author}{Tung, K.~K.} \& \bibinfo{author}{Orlando, W.~W.}
\newblock \bibinfo{title}{The ${k}^{\ensuremath{-}3}$ and
  ${k}^{\ensuremath{-}5/3}$ energy spectrum of atmospheric turbulence:
  Quasigeostrophic two-level model simulation}.
\newblock \emph{\bibinfo{journal}{J. Atmos. Sci.}}
  \textbf{\bibinfo{volume}{60}}, \bibinfo{pages}{824 -- 835}
  (\bibinfo{year}{2003}).

\bibitem{Gkioulekas/Tung:2006}
\bibinfo{author}{Gkioulekas, E.} \& \bibinfo{author}{Tung, K.-K.}
\newblock \bibinfo{title}{Recent developments in understanding two-dimensional
  turbulence and the nastrom--gage spectrum}.
\newblock \emph{\bibinfo{journal}{J. Low Temp. Phys.}}
  \textbf{\bibinfo{volume}{145}}, \bibinfo{pages}{25--57}
  (\bibinfo{year}{2006}).

\bibitem{Nastrom/etal:1984}
\bibinfo{author}{Nastrom, G.~D.}, \bibinfo{author}{Gage, K.~S.} \&
  \bibinfo{author}{Jasperson, W.~H.}
\newblock \bibinfo{title}{Kinetic energy spectrum of large-and mesoscale
  atmospheric processes}.
\newblock \emph{\bibinfo{journal}{Nature}} \textbf{\bibinfo{volume}{310}},
  \bibinfo{pages}{36--38} (\bibinfo{year}{1984}).

\end{thebibliography}

\end{document}